# High Electric Field Carrier Transport and Power Dissipation in Multilayer Black Phosphorus Field Effect Transistor with Dielectric Engineering

*Faisal Ahmed, Young Duck Kim, Min Sup Choi, Xiaochi Liu, Deshun Qu, Zheng Yang, Jiayang Hu, Irving P. Herman, James Hone, and Won Jong Yoo* [*]

F. Ahmed, Dr. M. S. Choi, X. Liu, D. Qu, Z. Yang, Prof. W. J. Yoo
Department of Nano Science and Technology
SKKU Advanced Institute of Nano-Technology (SAINT)
Sungkyunkwan University, 2066 Seobu-ro, Jangangu, Suwon, Gyeonggi-do, 440-746, Korea.
E-mail: yoowj@skku.edu

F. Ahmed, Prof. W. J. Yoo
School of Mechanical Engineering
Sungkyunkwan University, 2066 Seobu-ro, Jangangu, Suwon, Gyeonggi-do, 440-746, Korea.

Dr. Y. D. Kim, Prof. J. Hone
Department of Mechanical Engineering
Columbia University, 500 W. 120th St., New York, New York 10027, USA.

J. Hu, Prof. I. P. Herman
Department of Applied Physics and Applied mathematics
Columbia University, 500 W. 120th St., New York, New York 10027, USA.

**Abstract**

This study addresses high electric field transport in multilayer black phosphorus (BP) field effect transistors (FETs) with self-heating and thermal spreading by dielectric engineering. Interestingly, we found that multilayer BP device on a SiO$_2$ substrate exhibited a maximum current density of 3.3×10$^{10}$ A/m$^2$ at an electric field of 5.58 MV/m, several times higher than multilayer MoS$_2$. Our breakdown thermometry analysis revealed that self-heating was impeded along BP-dielectric interface, resulting in a thermal plateau inside the channel and eventual Joule breakdown. Using a size-dependent electro-thermal transport model, we extracted an interfacial thermal conductance of 1~10 MW/m$^2$·K for the BP-dielectric interfaces. By using hBN as a dielectric material for BP instead of thermally resistive SiO$_2$ ($\kappa$ ~ 1.4 W/m·K), we observed a 3 fold increase in breakdown power density and a relatively higher electric field endurance together with efficient and homogenous thermal spreading because hBN had superior structural and thermal compatibility with BP. We further confirmed our results based on micro-Raman spectroscopy and atomic force microscopy, and observed that BP devices on hBN exhibited centrally localized hotspots with a breakdown temperature of 600K, while the BP device on SiO$_2$ exhibited a hotspot in the vicinity of the electrode at 520K.

**1. Introduction**

In the pursuit of highly efficient low power miniaturized devices, novel two-dimensional (2D) layered materials have been explored over the last decade since conventional bulk materials have already been scaled to their geometrical dimension-performance threshold.[1-2] Interestingly, the characteristics of these layered 2D materials are dramatically different from their parent materials, especially when they are incorporated into solid state architectures. Atomically thin semiconducting 2D materials have numerous fascinating properties like mechanical flexibility, optical transparency, good electrostatic modulation and quantum confinement.[1-3] The thicknesses of these materials are smaller than their average phonon mean free path, which is ~50 to 300 nm near room temperature[2]. This has a number of effects including: (i) the formation of abrupt junctions in their immediate environment that result in inevitable and rather frequent phonon-boundary scattering,[2-3] and (ii) a significant reduction in the thermal conductivity ($\kappa$) due to a phonon confinement effect,[3] and (iii) improvements in packing density in integrated circuits and systems with increasing power dissipation density.[3,4]

Altogether, the aforementioned factors result in a substantial local temperature rise in functional devices and circuits based on 2D materials. Moreover, these devices are practically operated near current saturation conditions, *i.e.* under a high electric field, where charge carriers rigorously interact with each other, and also with phonons, material defects, impurities and sharp interfaces. Collectively, these scattering events further elevate the device operating temperature to a point where device breakdown occurs in a process called Joule breakdown. Heat removal is a formidable challenge that must be addressed to realize reliable operation of miniaturized devices based on novel 2D layered materials. Substantial efforts have been made under low electric field measurement conditions, but studies related to high electric fields and the corresponding power dissipation issue are rare. Therefore, we study power dissipation using high field breakdown thermometry for field effect transistor (FET) architectures using 2D black phosphorus (BP) as a channel material [5-7].

BP, a rare uni-elemental 2D material, has a sizeable, direct and thickness mediated optical band gap in the range of 0.3 to 2 eV, making it an ideal choice for numerous optoelectronic applications over a broad electromagnetic spectrum.[8] BP shows hole dominated ambipolar behavior with a hole mobility of ~1000 cm$^2$/V·s and current rectification in the range of $10^2 \sim 10^5$.[5–7] This addresses the shortcomings of other 2D materials like graphene and semiconducting transition metal dichalcogenides (TMDCs), which suffer from low on/off ratio and carrier mobility, respectively. Above all, BP exhibits pronounced in-plane directional anisotropy thanks to its puckered honeycomb lattice structure enabled by $sp^3$ hybridization between its orbitals.[6,7,9] For example, BP exhibits different in-plane $\kappa$ values along its zigzag ($\kappa_{zz} \sim$ 10 to 20 W/m·K) and armchair ($\kappa_{AC} \sim$ 20 to 40 W/m·K) directions depending on flake thickness.[9] However, the average $\kappa$ of BP ($\kappa_{avg} = \sqrt{\kappa_{AM} \times \kappa_{ZZ}} \approx$ 28.8 W/m·K) is smaller than that of graphene (< 2000 W/m·K[2]) and MoS$_2$ (85 W/m·K[10]) due to the large disparity in its in-plane phonon modes,[9] comparatively smaller contribution of the out-of-plane acoustic modes[11,12] and lower Debye temperature[12]. The smaller $\kappa$ value of BP together with its higher electrical conductivity ($\sigma$) make it a good thermoelectric material,[13] but results in impeded heat spreading during device operation.[2-4] Therefore, it is important from a device operation and reliability perspective to have a solid understanding of high-field transport and the corresponding power dissipation issues when trying to integrate BP into energy efficient electronic structures.

Previously, Engel *et al.* studied the power dissipation issues in BP using micro Raman techniques.[14] The scope of that paper was limited to measuring the local temperature rise during self-heating of BP flakes. In another study, the heat spreading in BP device is reported with the heat source being optical-absorption instead of Joule heating, to elucidate the thermally driven photocurrent generation.[15] In this study, we employed a simple and yet robust technique to elucidate thermal power dissipation together with efficient cooling of BP devices. We applied a high field breakdown technique to various BP FETs with different BP layer thicknesses. The applied electric field across the device was gradually increased to the point that the power deposited was large enough to cause breakdown. Our measurements showed that multilayer BP flake (11 nm) achieved a record high current level of 603 μA ($J_{max}$ = 3.3×10$^{10}$ A/m$^2$) at a maximum electric field of 5.58 MV/m. Surprisingly, the breakdown power scaled linearly with the footprint channel area ($L{\times}W$), which suggests that Joule heating in the channel was the likely breakdown mechanism. On the basis of this relationship, we deduced the interfacial thermal conductivity of 1~10 MW/m$^2$·K between BP-dielectric interfaces.[3,16] Furthermore, our findings indicate that the poor structural and thermal properties of conventional dielectric SiO$_2$ limit the heat dissipation during high field transport in BP devices. Employing hBN as the dielectric material instead of SiO$_2$ facilitated efficient and uniform heat dissipation mainly due to its higher in-plane $\kappa$ (~360 W/m·K[17]) and atomically clean surface. As a result, we observed a 3 fold increase in breakdown power density, a relatively higher electrical field endurance and a 13% increase in breakdown temperature for BP devices on a hBN substrate. This study provides important figures-of-merit and mechanisms that are crucial to improve the functionality and reliability of low power electronics especially under harsh environments.

## 2. Results and Discussion

The fabricated back gate BP device, shown in schematic diagram in Figure 1a and optical microscopy image in Figure 1b, was firstly characterized by applying a fixed gate bias ($V_G$) while sweeping the drain bias ($V_D$). Figure 1c shows the results obtained from a representative 11 nm thick BP device at different gating conditions. The linearity of the plots suggests Ohmic-like contact between Cr and BP. The current level at $V_G$ = 40 V was small, and it kept increasing as the applied $V_G$ decreased towards – 40 V. This trend confirmed p-type behavior of BP, as reported previously.[5-7] The bands tend to bend upwards as $V_G$ decreased, inducing smaller and narrower interfacial barriers for holes along the Cr-BP contacts. This resulted in an increase in current level, as indicated in the energy band diagram provided in the inset of Figure 1c. We also assembled a transfer plot of the same device, as shown in Figure 1d, which further confirms the p-type behavior with a hole current rectification ratio of ~$10^3$ and hole concentration of $n_{2D} = C_{ox} \times (V_G - V_{TH}) \approx 4.4 \times 10^{12}$ cm$^{-2}$ at $V_G = -40$V, where $C_{ox}$ is the capacitance per unit area to the back gate oxide ($C_{ox} = \varepsilon_\circ \varepsilon_r / t_{ox} = 1.2 \times 10^{-8}$ F/cm$^2$ for a 285 nm thick SiO$_2$) and $V_{TH}$ (19 V) is threshold voltage of BP device. The field effect mobility was extracted from a linear fit of the data in Figure. 1d using $\mu = g_m \frac{L}{W C_{OX} V_D}$. Here, $g_m$ is the trans-conductance ($\partial I_D / \partial V_G$), $L$ and $W$ are the channel length and width, respectively. Our device had dimensions of $L = 1.12$ μm and $W = 1.66$ μm. These values resulted in a hole mobility of 267 cm$^2$/V·s at $V_D = 0.1$ V under the room temperature. We understand that this value can be further enhanced by optimizing flake thickness [7] and employing a high-$k$ dielectric material [1].

After the low field electrical measurements, we next focused on higher electrical field ($V_D/L$) measurements to determine the sustainable electrical strength of BP. For these measurements, the electrical field applied to the multilayer BP device was continuously swept while gradually increasing the highest values unless electrical breakdown occurred. We observed a continual increase in current level with applied electrical field up to a certain maximum point, followed by a sudden drop in current level as shown in Figure 2a. The electrical breakdown occurred soon after reaching the maximum point, so the ultimate sustainable values of current, bias and electrical field were taken as the breakdown current ($I_{BD}$), breakdown voltage ($V_{BD}$) and breakdown field ($F_{BD}$), respectively. Using the 11nm thick BP device, we obtained an $I_{BD}$ of 603 μA ($J_{BD} = I_{BD}/W \times t = 3.3 \times 10^{10}$ A/m$^2$) at $V_{BD} = 6.25$V ($F_{BD} = 5.58$ MV/m). Generally, an increase in current level was observed as the applied field was increased, perhaps due to the increase in drift velocity of charged carriers and their corresponding reduction of the transit time ($time = L^2/\mu V_D$ ).[18] As the applied power increased, the device heated up to the extent that physical rupture, *i.e.*, Joule breakdown, occurred. The ultimate current carrying capacity of our ~11 nm thick BP FET, $3.3 \times 10^{10}$ A/m$^2$, is around seven times higher than the maximum reported value for multilayer MoS$_2$ in a similar geometry,[19] and was 3.3 times higher than the basic electron-migration limits for metals.[20,21]

In addition, we studied the thickness dependence of breakdown current in multilayer BP devices. In order to accomplish this, we fabricated various two terminal BP devices with different thicknesses (all having a ~1 μm long channel), and we measured their breakdown current ($I_{BD}/W$) at $V_G = 0$ as shown in Figure. 2b. Interestingly, among our studied devices, the highest current level of 666 μA/μm was recorded for the 41 nm thick sample. Previous results showed that thicker BP flakes exhibited higher $\kappa$ values and less surface scattering when compared to thinner flakes,[9] and this better explains the thermal spreading for thicker BP samples. We observed an increase in ultimate current level with increasing thickness. Note that $I_{BD}$ did not scale linearly with the thickness of BP flakes since the current distribution per layer was non-uniform in the multilayer BP, mainly due to charge screening and interlayer effects as observed for MoS$_2$ previously.[19,22] Afterwards, we studied the effect of lateral device dimensions ($L$ and $W$) on the electrical breakdown of BP. We initially fabricated devices with different $L$ values, which were fabricated on the same BP flake, as shown in the inset of Figure 2c, and we recorded their corresponding breakdown power ($P_{BD}$) (*i.e.*, the product of $I_{BD}$ and $V_{BD}$). Surprisingly, we found that the maximum power sustained by the BP FETs scaled linearly with $L$. However, a similar trend was observed for different $W$ devices as well [see the Supporting Information S1], indicating that $P_{BD}$ scaled linearly with the foot-print area ($L \times W$). Based on this, we initially assumed that the BP channel was subjected to Joule heating and that heat energy was spread-out along the in-plane and out-of-plane directions towards the BP-electrode (Cr/Au) and BP-dielectric ($SiO_2$) interfaces, respectively. Additionally, it seems that the former interface may serve as a more efficient heat sink than the latter interface due to better thermal coupling of metallic contacts with the BP flake compared to $SiO_2$. As a result, the edges of the channel cooled off, and the dissipated heat is trapped along the BP-$SiO_2$ interface, inducing thermal stresses inside the channel at a high electric field. This explains the linear trend between $P_{BD}$ and the channel cross-section. Further details about this understanding are provided below.

Owing to the 2D geometry of BP, the in-plane $\kappa$ value is larger than that of the out-of-plane value, mainly due to strong in-plane covalent bonds and weak van der Waals interactions along the c-axis, respectively, resulting in effective lateral thermal power propagation.[11] This seems to contradict our above speculation, where we assumed dominant heat spreading and trapping would occur in the out-of-plane direction. This contradiction suggests that $\kappa$ is not the only parameter that influences the heat dissipation direction, but device dimensions,[19] the nature of the interface [21] and surface conditions [2-4] may also affect the thermal spreading caused by Joule heating. From a geometry perspective, the out-of-plane cross-sectional area ($L \times W$) of BP devices is normally several orders larger than the in-plane cross-sectional area ($t \times W$), which results in a higher thermal conductance along the out-of-plane direction. Furthermore, at elevated lattice temperature, the higher frequency optical branches were excited, leading to an enhanced contribution of optical phonon branches and softening of flexural phonon branches (z-direction acoustic modes). This resulted in a suppressed $\kappa$ of BP, which in turn impeded the lateral heat propagation.[12] This observation suggests that the hot carriers were spatially confined near the center of the BP channel at high lattice temperature, inducing a temperature plateau (hot spots) inside the channel.[23] In short, due to centrally localized thermal carriers and the ultra-thin BP flake, the net thermal power dissipation occurs primarily towards the Si substrate. As mentioned previously, thicker BP flakes exhibited higher $\kappa$ values and less surface scattering when compared to thinner flakes.[9] Therefore, multilayer BP flakes with shorter and/or narrower channels would be desirable for effective heat spreading in operational electronic devices.

As mentioned earlier, the formation of thermally abrupt junctions masks the thermal transport in nanomaterials. Likewise, interfacial thermal properties greatly influence the operation of miniaturized devices and must be fully understood. Previous studies on 2D materials like graphene and $MoS_2$ supported on $SiO_2$ suggest that the oxide-channel interface is a bottleneck to the heat dissipation mainly due to weak thermal and structural coupling.[19,24] Special experimental setups were prepared in previous works to extract the interfacial thermal conductance ($G$).[10,25-27] However, in this study, we employed a highly robust analytical model based on electrical and thermal transport to extract $G$ per unit area of BP-dielectric interfaces.[3,16,28]

$$\begin{aligned} Q'' &= G\Delta T \\ P_o &= G(T_{BD} - T_o) \times A \end{aligned} \qquad (1)$$

$T_{BD}$ is the breakdown temperature of the BP FET and $T_o$ is room temperature. $Q''$ is the heat transfer per unit area, and $P_o$ is the breakdown power of the BP device excluding power dissipated along the contacts, *i.e.* $P_o = P_{BD} - I_{BD}^2 R_c$. Here, $R_c$ is the contact resistance of the BP device extracted using the transfer length method [see Supporting Information S2]. We obtained a $G$ of ~ 7.3 MW/m²·K for the BP-SiO$_2$ interface by linearly fitting the data in Figure 2c and using $T_{BD}$ ~ 520K for the BP FET on SiO$_2$. We prepared more than six different thickness BP devices, and extracted their corresponding $G$ values, which spanned in the range of 2 to 10 MW/m²·K. This variation is probably due to different BP-SiO$_2$ interface conditions, surface qualities and BP flake thicknesses. Similarly, we deduced $G$ for different BP-dielectric interfaces, as shown in Supporting Information S3. It seems that $G$ appears mainly dependent on nature of particular interface rather than thermal properties of dielectric material. A similar understanding was previously realized for other nano-materials and dielectric interfaces.[2,3,16] However, the extracted values were very close to the reported value for other 2D materials like MoS$_2$ and MoSe$_2$ on SiO$_2$, and they were around one order smaller than the reported values for graphene-SiO$_2$ interfaces.[10,25-27] Thermal decay length ($\lambda_{Th}$) of metal electrode is another parameter that indicates the dominant path of thermal power dissipation in a device. For example, if the channel is much longer than $\lambda_{Th}$, then heat will be mainly dissipated through the underlying substrate, while for comparatively equal or shorter channels it will predominantly sink through metallic electrodes.[24,28] $\lambda_{Th}$ is analogous to the electrical transfer length and can be extracted as

$$\lambda_{Th} = \sqrt{\kappa t / G} \qquad (2)$$

Here, $\kappa$ and $t$ are the in-plane thermal conductivity and thickness of the BP channel, respectively. Using $\kappa_{avg}$=28.8 W/m·K[9] and $t$ = 41 nm for our representative device, we extracted a value of $\lambda_{Th} \approx$ 400 nm. Our smallest channel is more than two times longer than $\lambda_{Th}$, and this further confirms that most of the power was vertically dissipated along the BP-SiO$_2$ interface. The readers should note that the above analytical model can only be used for devices with small $R_c$, but it may not hold well for semiconducting TMDCs that have a relatively large $R_c$.[29]

Owing to their high surface-to-volume ratio, 2D materials are highly sensitive to their immediate environment. Therefore, their electronic and phononic behavior can be easily altered by dielectric engineering. As explained above, the thermal energy is primarily transferred to the Si substrate through the dielectric during device operation. Likewise, devices fabricated on $SiO_2$ are subjected to thermal spreading problems due to its poor thermal conductivity ($\kappa \sim 1.4$ W/m·K) and corrugated surface.[21] Therefore, integration of thermally and structurally favorable dielectric materials instead of $SiO_2$ may greatly suppress these adverse effects and help keep the device cool during high field operation.

Hexagonal boron nitride (hBN) is a wide band gap (5.8 eV) layered dielectric material having a pristine flat surface, a good dielectric constant (~ 3.5), a high in-plane $\kappa$ (~ 360 W/m·K) and large surface optical phonon modes; these properties indicate that hBN is a strong candidate for use in 2D devices.[17,30] Previously, hBN was integrated with graphene and TMDCs for enhancing charge carrier transport.[30,31] Recently, hBN was integrated with BP as a capping layer[32] and an electrical performance booster for low field operations.[33,34] Therefore, we employed it as an alternative dielectric material for high field transport in BP devices. To this end, we exfoliated few-layer hBN flakes onto $SiO_2$ and mechanically stacked ~150 nm thick BP flakes over it using a dry transfer technique.[30,35] The BP flake was partially stacked over the hBN, as shown in Figure 3a, and two different devices with the same device dimensions were fabricated along the same in-plane direction of BP flake (*i.e.,* the zigzag direction) as confirmed by polarized Raman spectroscopy[36] to ensure a fair comparison. First, we characterized both devices at lower electrical fields as shown in Figure 3b. We did not observe any significant change in low field electrical characteristics for $SiO_2$ and hBN supported BP devices. We attributed the observed stubborn behavior of BP to the fact that the optical phonon scattering of BP in a low field may be the dominant scattering mechanism. This behavior may also be due to the weak charge screening effect caused by thicker BP flakes in our particular case. Previously, researchers reported that BP devices on hBN substrates showed a slight improvement in current level after a double annealing processing.[34] Based on this report, we believe that improvement can be partially attributed to annealing effects rather than the hBN dielectric alone. Afterwards, we slowly increased the applied electric field and, to our surprise, we observed an obvious change in higher electrical field transport for given devices. Our BP device on hBN exhibited a higher maximum power density and electric field sustainability than that on $SiO_2$, as shown in Figure. 3c. The BP device on $SiO_2$ exhibited an ultimate power of 33.25 mW at a maximum electric field of 2.25 MV/m, while a nearly 2-fold increase in power value (59.63 mW) and a comparatively larger field of 3.47 MV/m were realized on the hBN dielectric.

We repeated the experiment on more than 4 different devices and observe a 2 to 3 fold increase in maximum power values. These superior high field transport values were attributed to efficient thermal dissipation of BP devices on hBN compared to that on $SiO_2$. Structurally, hBN had an atomically flat and inert surface, while that of $SiO_2$ is highly corrugated and rough. Acoustic phonons, the dominant heat carriers in semiconducting materials, are more sensitive to interface scattering than their optical companions.[2] As such, the rough surface of $SiO_2$ may greatly limit the thermal transport in the device, whereas relatively smooth heat conduction can be obtained using hBN. Additionally, hBN has a ~250-fold higher $\kappa$ and 2-fold higher surface optical phonon energy compared to $SiO_2$, which enabled relatively better thermal coupling of hBN to BP. This further facilitated thermal spreading during high field operation.

To quantitatively analyze the impact of dielectric engineering on the high field transport of the BP device, we performed micro Raman spectroscopy to extract the local temperature increase in the device. Micro Raman spectroscopy is a non-invasive approach for determining the phonon temperature, and it has previously been employed in 2D materials like graphene [23,37] and BP[14]. Further details about our micro Raman setup can be found in Ref. 23. The crystalline multilayer BP exhibited three dominant Raman peaks. The two in-plane modes of $A^2g$ and $B_2g$ represent atomic oscillations along the zigzag and armchair directions, respectively, and one out-of-plane mode, $A^1g$, depicts the *z*-direction lattice vibration.[14,36,37] We performed Stokes (positive) and anti-Stokes (negative) Raman spectroscopy on our BP devices on $SiO_2$ and hBN dielectrics under an applied voltage as shown in Figures 4a and 4b respectively. The measured zero bias ($V_D$ = 0V) Raman peaks at ± 365, ± 442, ± 470 cm$^{-1}$ were attributed to the corresponding $A^1g$, $B_2g$, and $A^2g$ phonon modes of crystalline BP. Further, we gradually increased the applied bias and recorded the Raman signal so as to observe Raman peak softening with increasing $V_D$ for BP devices on both the dielectric materials. It should be noted that no gate bias was applied during Raman measurements since the thicker BP flakes usually showed immunity towards gating mainly due to weak charge screening. Thus, the spectral shift in Raman spectra is solely caused by electrical heating of BP lattice. Generally, the intrinsic softening of Raman peaks due to increase in flake temperature is attributed to the thermal expansion of lattice and an-harmonic phonon coupling.[36] Therefore, a clear red shift in Raman spectra of BP lattice, as shown in Figures 4a and 4b, is mainly realized due to self-heating of BP flake by applied electrical bias. It is important to note that the similar shift in Raman modes was realized by direct thermal heating of BP flake as well.[36] The further details about spectral peaks position shift and related extraction of electrical heating coefficients of specific Raman modes of multilayer BP are given in Supporting Information S4. The deconvoluted intensity ratios of Stokes ($I_S$) and anti-Stokes ($I_{AS}$) peaks were plotted as a function of applied electrical power as shown in Figure 4c and 4d on SiO$_2$ and hBN substrates, respectively. As shown, we observed a linearly increasing trend for all the three Raman modes. This ratio can be translated to a temperature by using Equation (3).[23]

$$\frac{I_{AS}}{I_S} \alpha \ C \exp\left(-\frac{E_{op}}{k_B T_{ph}}\right) \qquad (3)$$

Here, $T_{ph}$ is the phonon temperature, $E_{op}$ is the optical phonon energy of each Raman peak: $A^1g$ = 45.19 meV, $B_2g$ = 54.68 meV, and $A^2g$ = 58.15 meV. $C$ is the measured pre-factor due to the CCD response and optics, which were carefully calibrated. We obtained an operating temperature at a given applied power value for the $A^2g$ mode by using the ratios of Stokes and anti-Stokes intensities from Figure 4c and 4d in Equation (3), as shown in Figure 4e. In addition to this, we employed an analytical model based on heat diffusion equation to compute the operating temperature [See Supporting Information S5 and S6]. The obtained results are shown in Figure 4e by solid lines, and they fit well with our experimentally determined temperature numbers. However, from analytical and experimental temperature results we observed that BP on hBN showed relatively lower operating temperatures (*i.e.* cooler device operation) than that of SiO$_2$ under the same applied field conditions. This indicates efficient heat dissipation in the hBN supported BP device.

Similarly, we obtained peak operating temperatures of 520 K and 600 K at the breakdown point for the BP device on SiO$_2$ and hBN substrates, respectively. SiO$_2$ is known for enhancing surface scattering, mainly due to surface polar optical phonon scattering *via* remote-phonon interactions and charged impurity scattering, which cause hot carrier relaxation and eventually affect the local temperature in the device.[23]. In contrast, the atomic level flatness of hBN enables intimate thermal contact with BP, causing better phonon-phonon interactions between them, which ensure relatively cooler device operation. This observation further confirmed that BP on hBN can withstand a higher maximum power density and operating temperature due to efficient cooling of a device at high field operation. Our observed breakdown temperature values on both substrates were smaller than previously extracted for BP, *i.e.,* 757K in ref. 14. We think this difference may be due to different quality and thicknesses of BP used, different processing and operating conditions adopted and more notably different dielectrics than were previously used (200 nm ITO and 100 nm $Al_2O_3$). Nonetheless, at such high temperature, the crystalline black phosphorus flake may have already changed to amorphous red phosphorus.[39]

Finally, we inspected the devices after electrical breakdown under an optical microscope. Interestingly, we observed that the BP device on $SiO_2$ experienced cracks in the vicinity of the electrode, while it is located along the center of the channel for hBN as shown in Figure 5a. We further confirmed this anomaly in BP devices on $SiO_2$ by using AFM, as shown in Figure. 5b. The observed crack position from the AFM image indicates that a hot spot was induced ~450 nm away from the metal electrode, which is also consistent with the thermal decay length of metal electrode ($\lambda_{Th} \approx$ 400 nm). Based on the location of thermally induced cracks, we speculated that thermal spreading was non-uniform for BP devices on $SiO_2$, whereas it seemed more homogeneous on a hBN substrate.

The position of hot spots on BP devices on $SiO_2$ is also of interest. It was previously reported that the thermally induced trapped charges in $SiO_2$ resulted in an abrupt doping profile below the 2D material.[40-42] Therefore, a hot spot was induced near regions of low carrier density (biased contact). This may be the reason that we observed thermally induced cracking near the metallic electrode (drain) on SiO$_2$. Based on above discussion, it is clear that the poor structural and thermal properties of SiO$_2$ not only impeded thermal distribution that masked device operating temperature, but also resulted in uneven heat spreading that caused rupture near the metallic electrode. On the other hand, better structural and thermal coupling of hBN with BP helped realize homogenous thermal spreading while enabling cooler device operation. This allowed us to achieve centrally localized hotspots and relatively higher sustainability of breakdown power density compared to devices on SiO$_2$. In the future, this inhomogeneity can be further studied in further detail by employing spatial resolution techniques, *e.g.* scanning thermal microscopy, infrared spectroscopy[43] and scanning Joule microscopy[44]. Our analysis demonstrated that hBN effectively protected BP from environmental perturbations and improved performance under low fields, making it a favorable dielectric material for high field operation.

## 3. Conclusion

In conclusion, we applied breakdown thermometry to study the power dissipation in BP FETs. We found that multilayer BP exhibited a higher current density than that of multilayer MoS$_2$ in a back gate device structure. Moreover, the interfacial thermal conductance between BP-dielectric interface was extracted by implementing a simple analytical approach. Finally, the dielectric material greatly influenced high field operation. Similarly, efficient device cooling was achieved by employing hBN as a dielectric for BP devices instead of SiO$_2$.

## 4. Experimental Section

*Device fabrication*: Multi-layer BP flakes were placed on a p-doped Si substrate capped with thermally grown 285nm $SiO_2$ in an Ar atmospheric glove box having oxygen and moisture levels < 1 ppm. Candidate flakes were targeted by optical contrast, and an electron beam resist polymer (PMMA) was coated on the substrates inside an environmentally controlled glove box. Electrodes were patterned *via* electron beam lithography (EBL) and 5/50 nm thick Cr/Au metal layers were deposited by electron beam deposition (EBD) followed by lift-off in acetone to remove excessively deposited metal. A schematic of the simple two terminal back gate BP FET device is shown in Figure 1a, and an optical microscopy (OM) image of an ~11 nm thick BP device is shown in Figure 1b. The thickness of the BP flakes was measured using atomic force microscopy (AFM). An error of ±1 nm is appropriate for these measurements due to the collection of moisture over the BP surface. During the fabrication process, extra efforts were made to minimize BP exposure to the ambient environment to ensure high quality BP devices. Soon after lift-off, electrical measurements were carried out in a vacuum environment, and our measurements lead to subsequent breakdown of BP devices. Therefore, the total environmental exposure was very short, hence the probability of oxidation of our BP devices is very low. We recently studied the stability and effective passivation of BP flakes elsewhere.[45]

*Micro Raman Spectroscopy*: Micro Raman spectroscopy were acquired using the 514.5 nm Ar laser with a power of 300 μW and spot size of 1 μm under vacuum (~ $10^{-5}$ torr) with applied electric field to multilayer BP devices. We used a long working distance ×50 object lens (Olympus LMPLFN50x) and spectrometer (Princeton Instrument, eXcelon-100B, 1,800 groove/mm grating) with 30 sec exposure time.

**Acknowledgements**

This work was supported by the Global Research Laboratory (GRL) Program (2016K1A1A2912707) and the Global Frontier R&D Program (2013M3A6B1078873) at the Center for Hybrid Interface Materials (HIM), funded by the Ministry of Science, ICT & Future Planning via the National Research Foundation of Korea (NRF). Y.D.K and J.H. were supported by grant from the ONR (N00014-13-1-0662 and N00014-13-1-0464) and DE-SC0012592.

# Figures

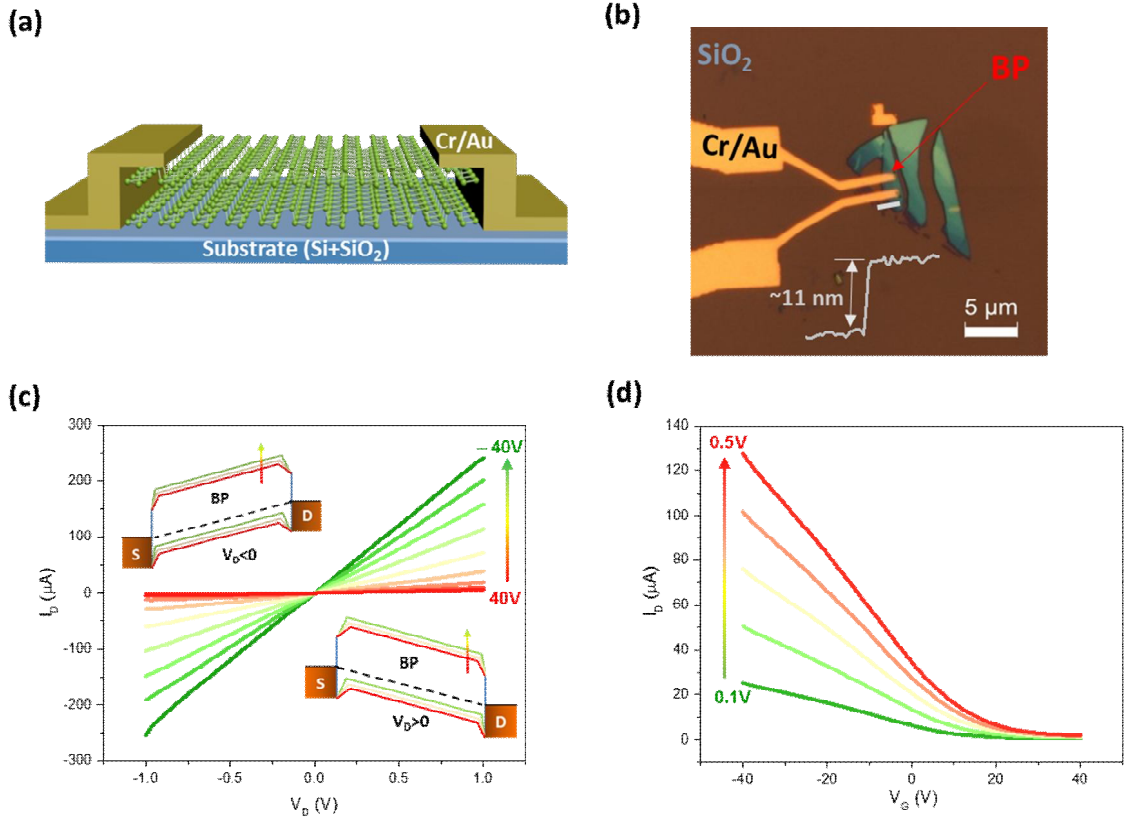

**Figure 1. Low field electrical characterization of multilayer BP device. a)** Schematic of a simple two terminal back gate BP FET device. **b)** Optical microscope image of a representative multilayer BP device, where inset shows the AFM thickness profile along the given white line on BP flake having ~11nm thickness. **c)** Output curve of device shown in b at a different gate biases with a step of 10V. Inset denotes the energy band diagram at different applied bias conditions. The color code indicates different gating conditions, while the upper left and lower right diagrams represent the band position at negative and positive $V_D$ conditions, respectively. **d)** Transfer curve at various drain biases (0.1 V steps).

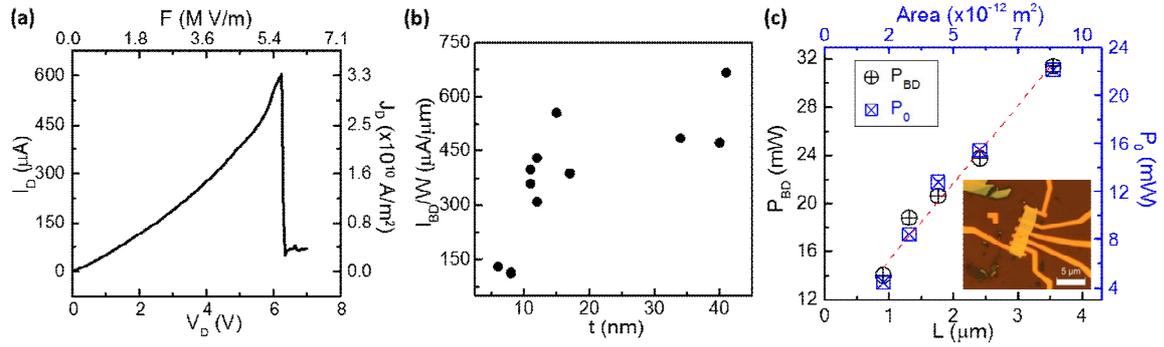

**Figure 2. Electrical breakdown of BP and its dependency on device dimensions. a)** The current density of the back gate BP FET device at zero gate bias plotted against applied bias and electrical field. **b)** Thickness dependent maximum current level of various ~1 μm long BP-FETs. **c)** Breakdown power ($P_{BD}$ and $P_0 = P_{BD} - I_{BD}^2 R_c$) obtained from different channel length devices fabricated on the same BP flake. The top axis denotes the corresponding cross-sectional area ($L \times W$). The inset shows an OM image of the device.

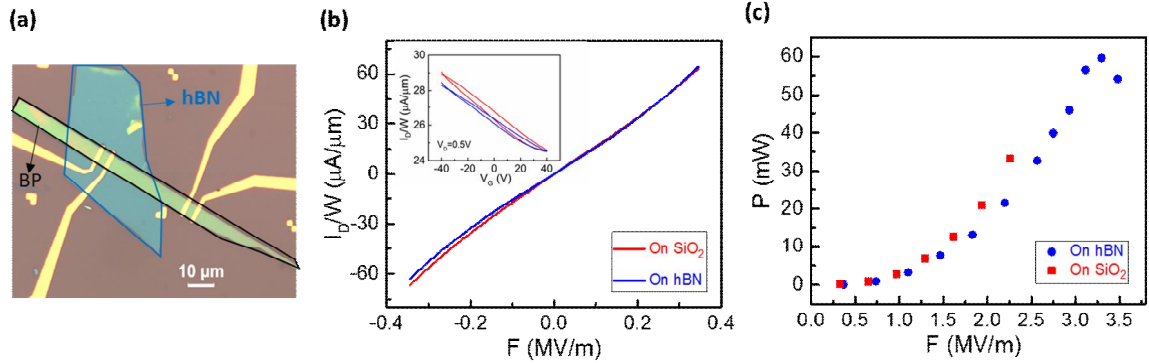

**Figure 3. Dielectric engineering to multilayer BP device. a)** Optical microscope image of BP flake partly stacked over hBN. The black and blue bordered areas indicate ~150nm thick BP and 15nm thick hBN regions, respectively. **b)** Low field electrical transport behavior of BP device on $SiO_2$ and hBN at $V_G$=0. The inset shows their corresponding hysteresis plots at $V_D$=0.5V. **c)** The obtained electrical power plotted against an applied electrical field for a BP device on $SiO_2$ and hBN.

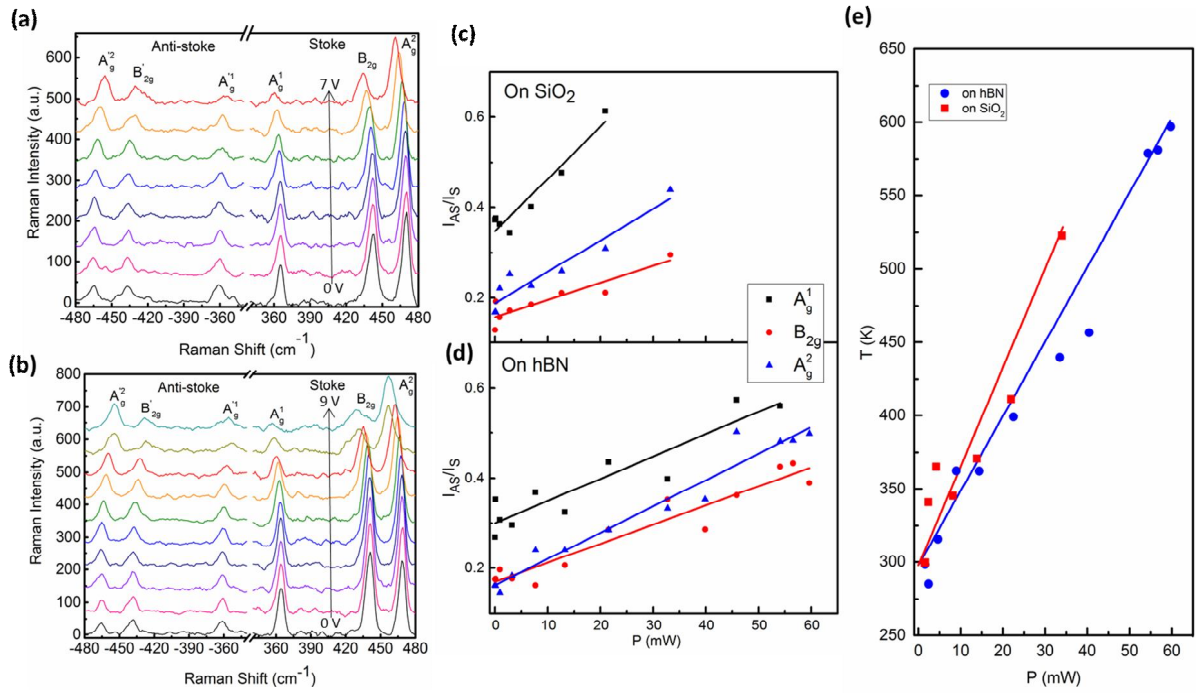

**Figure 4. Temperature extraction from micro-Raman spectroscopy**. **a)** and **b)** The obtained Stokes and anti-Stokes Raman spectra of BP device on $SiO_2$ and hBN, respectively, at different applied bias conditions. **c)** and **d)** represent the ratio of deconvolution Stokes and Anti-Stokes Raman peaks from (a) and (b), respectively, plotted as a function of applied power. **e)** The calculated phonon temperature of BP device on $SiO_2$ and hBN at a given power density. The lines represent analytically computed temperatures and the solid points are the experimentally determined temperatures from micro-Raman signals. Note that this temperature is extracted by comparing the $A^2_g$ phonon modes.

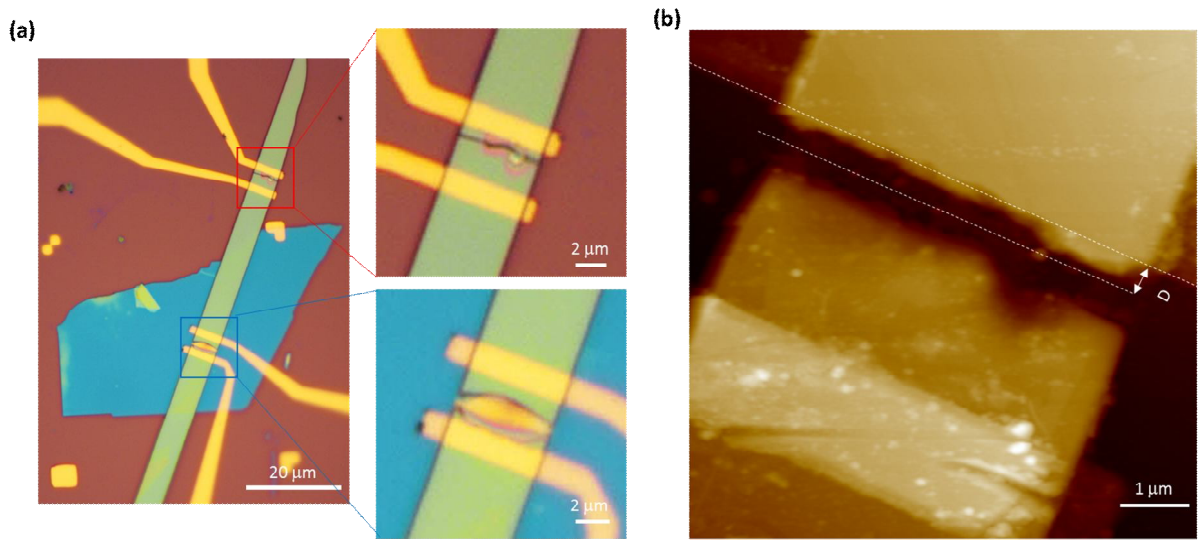

**Figure 5. Optiacal and atomic force microsopy of BP devices after electrical breakdown. a)** Optical microscope image of BP devices on SiO$_2$ and hBN substrate after electrical breakdown. The red and blue colored squares indicate the SiO$_2$ and hBN supported BP devices respectively **b)** AFM image of ~150nm thick broken BP device on SiO$_2$ substrate, where D denotes the distance between the electrode and hot-spot.

Supporting Information for

**High Electric Field Carrier Transport and Power Dissipation in Multilayer Black Phosphorus Field Effect Transistor with Dielectric Engineering**

*Faisal Ahmed, Young Duck Kim, Min Sup Choi, Xiaochi Liu, Deshun Qu, Zheng Yang, Jiayang Hu, Irving P. Herman, James Hone, and Won Jong Yoo* [*]

F. Ahmed, Dr. M. S. Choi, X. Liu, D. Qu, Z. Yang, Prof. W. J. Yoo
Department of Nano Science and Technology
SKKU Advanced Institute of Nano-Technology (SAINT)
Sungkyunkwan University, 2066, Seobu-ro, Jangangu, Suwon, Gyeonggi-do, 440-746, Korea.
E-mail: yoowj@skku.edu

F. Ahmed, Prof. W. J. Yoo
School of Mechanical Engineering
Sungkyunkwan University, 2066, Seobu-ro, Jangangu, Suwon, Gyeonggi-do, 440-746, Korea.

Dr. Y. D. Kim, Prof. J. Hone
Department of Mechanical Engineering,
Columbia University, New York, New York 10027, USA.

J. Hu, Prof. I. P. Herman
Department of Applied Physics and Applied mathematics,
Columbia University, New York, New York 10027, USA.

**Supporting Information**

1. **Relationship between breakdown power and channel width**

2. **Extraction of contact resistance**

3. **Interfacial thermal conductance of BP-dielectric interfaces**

4. **Calculation of electrical heating coefficients**

5. **Analytical extraction of temperature distribution at high electric field**

6. **Analytical calculation of operating device temperature as a function of applied electrical power**

## S1 (Relationship between breakdown power and channel width)

We fabricated BP devices with different channel widths ($W$) to ascertain their relationship with the breakdown power ($P_{BD}$). Towards this end, devices with three different widths, *i.e.* 3.67μm, 5.9μm and 6.75μm were fabricated on the same BP flake, all having ~1 μm channel length ($L$). As explained in the main text, we conduced study on the electrical breakdown and recorded their $P_{BD}$ as indicated in Figure S1a. Similar to the case of different $L$, $P_{BD}$ scaled linearly to $W$ and this result convinced that $P_{BD}$ depends linearly on foot-print area ($L \times W$) rather than $L$ or $W$ individually [see Figure S1b].

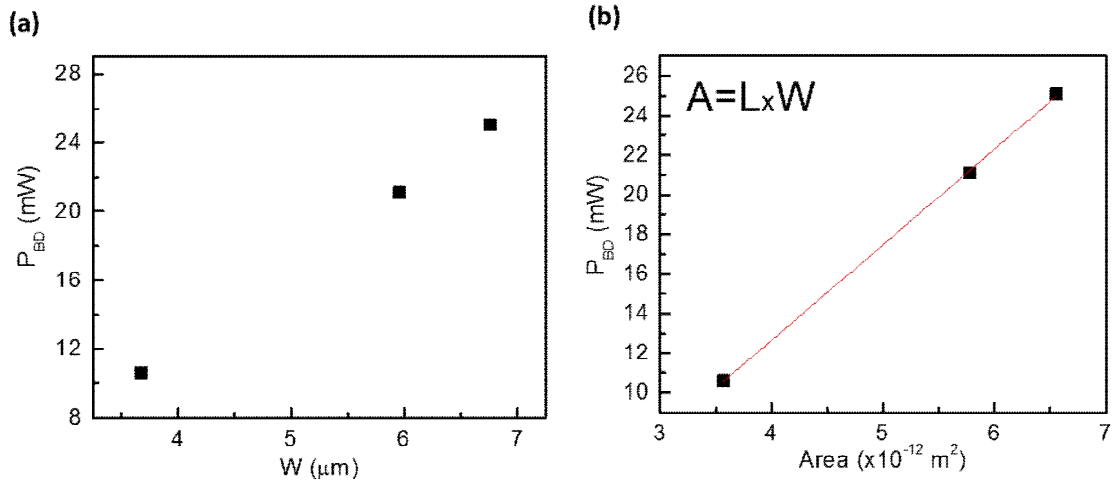

**Figure S1. Breakdown power *vs.* channel width. a)** The measured $P_{BD}$ at $V_G = 0$ from BP devices having same channel length and with three different widths. **b)** $P_{BD}$ plotted agianst foot-print area.

## S2 (Extraction of contact resistance)

We extracted contact resistance ($R_c$) of BP device by using transfer length method (TLM).[1] The low field transfer curves of devices with different $L$ are given in Figure S2a, obtained from BP device shown in the inset of Figure 2c in the main text. All the given BP devices shows dominant p-type behavior, however the longer channel ($L_5$) exhibits higher hole current level which increases by reducing $L$ towards shorter channels *i.e.* from $L_5$ to $L_1$. Moreover, we extracted the total resistance ($R$) of all five devices at the higher applied electric field (near breakdown point), and the result showed linear trend to the $L$ as shown in Figure S2b and thereby lineally extrapolating the plot to the y-axis, we obtained $R_c$ of 700 Ω at $V_G = 0$.

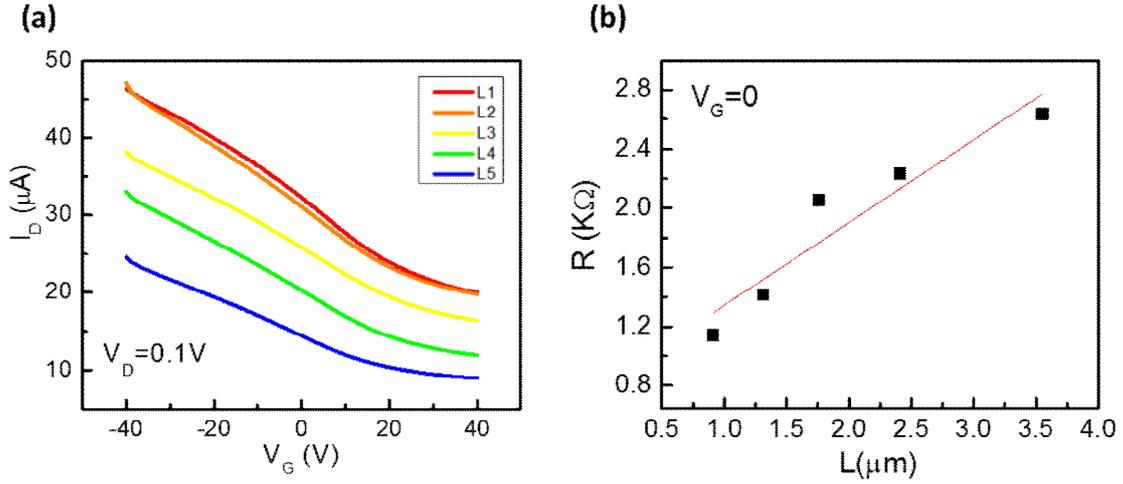

**Figure S2. Extraction of contact resistance. a)** The transfer curves of BP devices with different $L$ at $V_D = 0.1$ V. **b)** Total resistance of the BP devices at high field condition and $V_G = 0$. The black squares indicate measured data points.

**S3 (Interfacial thermal conductance of BP-dielectric interfaces)**

We prepared two different kinds of BP-dielectric interfaces, that is, BP-SiO$_2$ and BP-hBN and deduced their interfacial thermal conductance ($G$) values. For this, firstly BP devices with six different thicknesses were fabricated on SiO$_2$ substrate, and their $G$ value ranges from 2 to 10 M W/m$^2$·K. Thereafter, we fabricated BP devices with two different thicknesses on hBN substrate and similarly using size-dependent analytical model, we computed their $G$ values and the obtained results spanned in the range of 3 to 5 M W/m$^2$·K. The difference between the values of the two interfaces may be due to different condition of their particular interface and more importantly the different thickness of BP and hBN used. It is important to note that the average $G$ value of the BP-hBN interface is smaller than that of the BP-SiO$_2$ interface, and we think this may be due to the thermal healing effect of electrode in the latter case. The breakdown position of BP on $SiO_2$ substrate was always located in the vicinity of electrodes, as shown in Figure 5 of the main text. In that case, the heat may dominantly sink through electrodes. Therefore, we think the large average G value of BP on $SiO_2$ can be obtained due to the contribution of metal electrodes. More importantly, this also shows that $G$ may not be limited by thermal properties of dielectric itself.

**S4 (Calculation of electrical heating coefficients)**

The temperature dependent Raman shift can be used to define the vibration properties such as electro-phonon phonon-phonon coupling, or thermal expansion of materials.[2,3] Similarly, based on high field transport induced self-heating coupled with micro-Raman processing, it is possible to extract the electrical heating coefficients of Raman modes of multilayer BP.

For this, we deconvoluted the Stokes mode Raman intensities of multilayer BP on $SiO_2$ and hBN substrates as a function of applied electric bias from Figure 4a and 4b of the main manuscript, from which we found that all three spectral peak positions *i.e.* $A^1_g$, $B_{2g}$ and $A^2_g$ of multilayer BP supported on $SiO_2$ and hBN substrates showed a linearly decreasing trend, as depicted in Figure S4a and S4b.

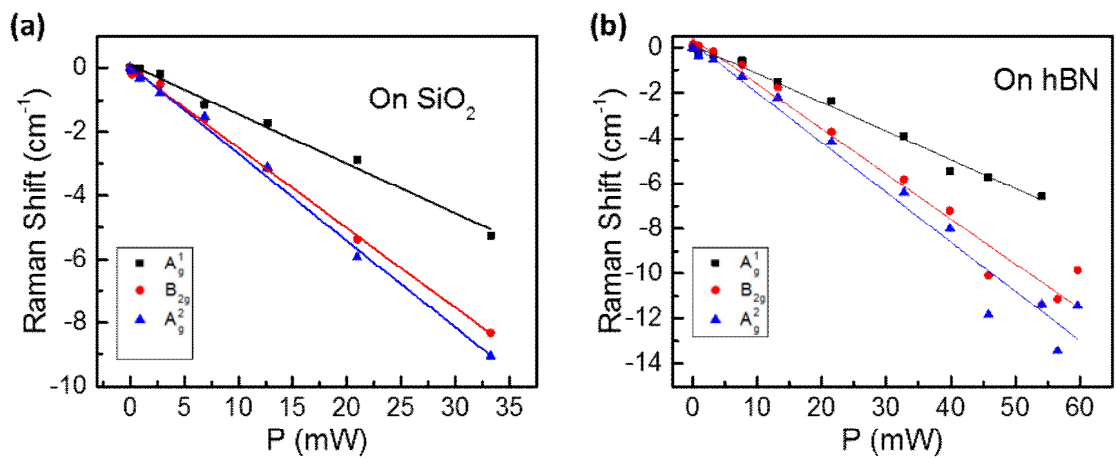

**Figure S4. Electrical heating coefficients a)** and **b)** The spectral Raman shift of multilayer BP flake as a fucntion of applied electrical power on SiO$_2$ and hBN substrate respectively. The lines indicate the linear fits to the data points of specific Raman modes.

From that, we obtained the slopes (Δ) of the specific Raman peak by linearly fitting its data to the applied electric power and the results are listed in Table S3.

| BP device | Δ (cm$^{-1}$/mW) | | | g (cm$^{-1}$/K) | | |
|---|---|---|---|---|---|---|
| | A$^1$g | B$_2$g | A$^2$g | A$^1$g | B$_2$g | A$^2$g |
| On SiO$_2$ (this work) | 0.154±0.005 | 0.251±0.004 | 0.273±0.006 | 0.026 | 0.043 | 0.047 |
| On hBN (this work) | 0.126±0.003 | 0.198±0.009 | 0.219±0.011 | 0.024 | 0.038 | 0.042 |
| On Al$_2$O$_3$ (ref. 5) | 0.18±0.05 | 0.44±0.05 | 0.5±0.05 | 0.013 | 0.033 | 0.038 |

**Table S3.** Comparison of slopes (*Δ*) and electrical heating coefficients (*g*) of specific Raman modes of BP on different substrates.

Subsequently, based on the temperature results of Figure 4e in the main manuscript, we scale the Raman shift of each mode to compute the electrical heating coefficients (*g*) for multilayer BP flake on both substrates. The obtained slope and coefficient values, as shown in Table S4, are very close to the previously reported numbers by electrical heating of multilayer BP on Al$_2$O$_3$ substrate,[4] and they are also on the same order of magnitude to the recently reported thermal heating coefficients of multilayer BP.[2,3]

It is noteworthy that the extracted Δ and *g* values of hBN supported BP flake are smaller than those of SiO$_2$ supported among our results. It is the matter of fact that the peak shift is readily dependent on thermal expansion coefficient (TEC) mismatch of the given materials.[3] Therefore, we think this is due to difference in TEC mismatch between BP-SiO$_2$ and BP-hBN interfaces, or the large Raman peak shift of BP on SiO$_2$ substrate may be attributed to the dominant thermal expansion caused by large temperature gradient due to non-homogeneous thermal spreading. However, further studies are needed to address this difference.

**S5 (Analytical extraction of temperature distribution at high electric field)**

We employed an analytical model based on heat diffusion equation to extract the temperature distribution for BP device near breakdown point.[5] Here, we assume that thermal conductivity of BP is independent of position and temperature of the device and the applied electrical power is homogeneously distributed along the BP flake on SiO$_2$ and hBN substrates. The heat diffusion equation is,

$$\frac{d^2 T}{dx^2} + \frac{1}{\kappa t}\left[\frac{P}{LW} - 2G(T - T_o)\right] = 0, \qquad (S1)$$

Here, $x$ is position along the channel. Now, solving S1 to extract maximum operating temperature as a function of location along the channel *i.e. T(x)* will yield,

$$T(x) = T_o + \frac{P}{c^2 \kappa LWt}\left(1 - \frac{\cosh(cx)}{\cosh(cL/2)}\right), \qquad (S2)$$

where $c = \sqrt{2G/kt}$. By applying corresponding values and breakdown power values of 33.25 mW and 59.63 mW for BP on SiO$_2$ and hBN respectively, we can extract the temperature distribution at breakdown point along the BP devices. It should be noted that our calculated $G$ values for BP on SiO$_2$ and hBN substrates span from 2 to 10 M W/m$^2$·K and 3 to 5 M W/m$^2$·K respectively: see Supporting Information S3. However, in this case the best temperature results that were in agreement with experimentally calculated values, were obtained at $G$ values of 3 M W/m$^2$·K and 5.1 M W/m$^2$·K for BP-SiO$_2$ and BP-hBN interfaces respectively, as shown in Figure S5a and S5b.

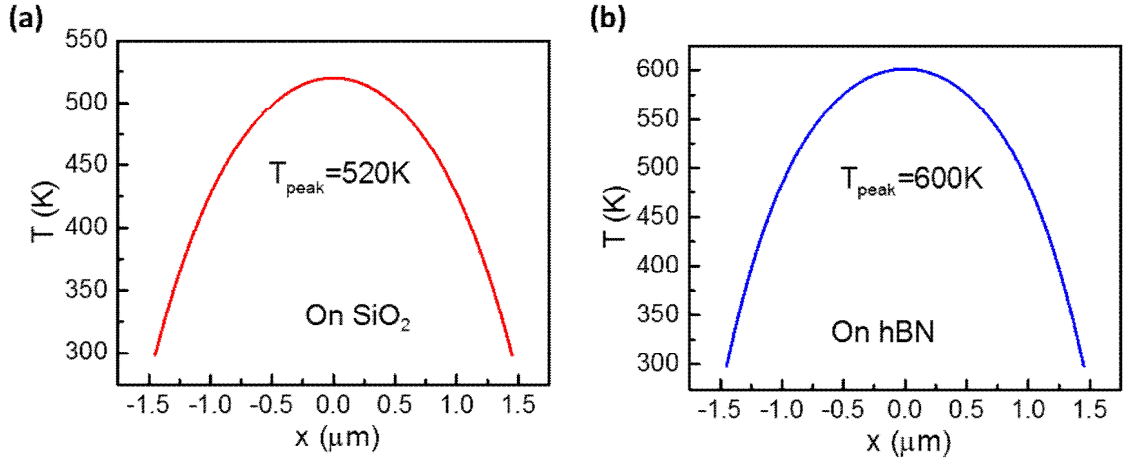

**Figure S5. Temperature distribution at high electric field a)** and **b)** The obtained temperature profiles of SiO$_2$ and hBN supported BP deviecs, with peak temperatures of 520 K and 600 K respectively.

Nonetheless, at all the given *G* values, the obtained temperature profiles were similar *i.e.* dome shaped plot while they differ in peak temperature values. Furthermore this profile indicates that, the center is heated, while contacts remain at room temperature. The readers may argue over the peak temperature position on SiO$_2$ substrate, since the hotspot location was near to the electrode from the optical microsope image [see Figure 5b in the main manuscript]. This contradiction is due to the fact that, in given model, we assume uniform power distribution. This model is more realistic for hBN supported or suspended devices since charge trapping at the dielectric is minimum or absent, which enables uniform power distribution along the BP device.

**S6 (Analytical calculation of operating device temperature as a function of applied electrical power)**

Next, we also computed operating temperature as a function of applied power by modifying equation S2. Since the operating temperature is maximum at the center of the flake, therefore setting *x* = 0 in S2,

$$T(P) = T_\mathrm{o} + \frac{P}{c^2 \kappa LWt}\left(1 - \frac{1}{\cosh(cL/2)}\right), \qquad (S3)$$

Similarly, using *G* values of 3 M W/m$^2$·K and 5.1 M W/m$^2$·K for BP-SiO$_2$ and BP-hBN interfaces respectively, we extracted temperature as a function of applied power for both the cases, as shown by solid line in Figure 4e of the main manuscript. The computed results coincide well with experimentally calculated temperature values.